\documentclass[dvips]{article}

\usepackage{icrc2011}

\title{A newly discovered VHE $\gamma$-ray PWN candidate around PSR J1459-60}

\newcommand{\etal}{\MakeLowercase{\textit{et al. }}} 
\shorttitle{R. de los Reyes \etal VHE $\gamma$-ray PWN candidate around PSR J1459-60}

\authors{R. de los Reyes$^{1}$, A. Zajczyk$^{2}$, R.C.G. Chaves$^{1}$ on behalf of the H.E.S.S. Collaboration}
\afiliations{$^1$Max-Planck-Institut f\"ur Kernphysik, PO Box 103980, 69029 Heidelberg, Germany\\ 
  $^2$Nicolaus Copernicus Astronomical Center, Warsaw, Poland}
\email{reyes@mpi-hd.mpg.de}

\abstract{Observations of the Galactic Plane performed by the H.E.S.S. telescope array have revealed a significant excess at very-high-energies (VHE; E$>$0.1 TeV) from the direction of PSR J1459-60, a rather old gamma-ray pulsar (64 kyr) with a spindown energy of $\sim$10$^{36}$ erg/s, discovered by the {\it Fermi}/LAT satellite in high-energy (HE) $\gamma$-rays. The X-ray pulsar counterpart has been recently detected using the {\it Suzaku} satellite.

In this contribution, we present the discovery of a new VHE $\gamma$-ray source, including morphological and spectral analyses. Its association with the $\gamma$-ray pulsar in a PWN scenario will be discussed.}

\begin{document}
\maketitle
\section{Introduction}

In the last decades, the population of Galactic sources at VHE $\gamma$-rays has increased considerably proving the domain of the Imaging Atmospheric Cherenkov Telescope (IACT) technique in this energy range.
Among the TeV galactic population, the majority of them are supernova remnants (SNRs) and pulsar wind nebulae (PWNe), although a large number of Galactic sources is still unidentified without counterpart at any other wavelength.
The detection and study of SNRs and PWNe have revealed them as efficient accelerators of cosmic rays (up to $\sim$100 TeV) in our Galaxy. However, to understand the acceleration mechanism inside these sources observations at different wavelengths, specially from X-rays to TeV $\gamma$-rays, are required. 
In particular, gamma-rays above 100 MeV will be a direct probe of the acceleration of the cosmic rays to ultra relativistic energies. These high energy particles produce $\gamma$-rays by two different mechanisms: through Inverse Compton (IC) scattering of the electrons with photon fields (like cosmic microwave background radiation (CMBR), star light, infrared (IR) dust emission or produced synchrotron radiation) or through  $\pi$$^0$s decay from the interaction of protons with interstellar material or close molecular clouds.

In the VHE domain, the H.E.S.S. Galactic Scan has contributed significantly to increase the population of sources. Up to now 5 TeV shell type SNRs and more than 26 PWNe have been detected by H.E.S.S.

In this contribution we present the discovery of HESS J1458-608 at TeV energies, a PWN candidate around one of the {\it Fermi}/LAT discovered $\gamma$-ray pulsars, PSR J1459-60. After the description of the observations and results of the data analysis, arguments about a possible PWN scenario will be discussed.

\section{Observation and data analysis}\label{sec:analysis}

The source presented here was discovered in the first place in the Galactic Plane Survey, performed by H.E.S.S. since 2004. After its first marginal detection (2$\sigma$ post-trials) dedicated observations were carried out in 2009 and 2010. Since the field of view (FoV) of the region around this source (l=317.75$^\circ$, b=-1.7$^\circ$) is quite crowded at VHE, the analysis data-set presented in this work includes an important fraction of observations of nearby sources.

After the standard H.E.S.S. quality cuts on the observation runs~\cite{QualityCuts}, we get 96 hours of total live time. Since the data set includes observations of nearby sources (at offset angles from the source position $>$1.5$^\circ$), the mean offset is 1.8$^\circ$, while the dedicated observations are taken at typical offset between 0.7$^\circ$-1.0$^\circ$. The mean zenith angle of this data set is 39$^\circ$, approximately the culmination zenith angle of the source. 

The results were obtained using the standard H.E.S.S. analysis~\cite{QualityCuts} based on cuts on the Hillas second moment parameters~\cite{Hillas}. {\it Hard cuts} and {\it TMVA} cuts~\cite{TMVA} have been applied to remove hadron $\gamma$-like events from the $\gamma$-ray events candidates.
These analysis cuts include a cut in the number of photo-electrons recorded from the Extended Air Shower (EAS) ($>$ 160 p.e.) and several others that are optimized to get the best angular resolution ($\sigma_{68\%}$=0.09$^\circ$)~\cite{Model++} and suppress the contamination from the nearby sources. A most sensitive analysis at high energies is also achieved despite a high energy threshold of 0.75 TeV, also due to the zenith angle range of the observations.

To obtain the sky maps, the background~\cite{BgMethods} was estimated for each bin from a ring of inner radius of 0.9$^\circ$ covering an area 10 times larger than the on region. For the spectral analysis the Reflected background method was used. This method integrates the signal from regions with the same size as the on region but placed at an offset of 0.4$^\circ$ from the center of the FoV.
All the presented results are consistent with a complete alternative analysis chain which includes an independent calibration and $\gamma$/hadron separation method~\cite{Model++}.

\section{Results}\label{sec:Results}

After camera acceptance correction w.r.t. the $\gamma$-ray pulsar coordinates (RA=14$^h$59$^m$29.8$^s$, Dec=-60$^\circ$52'40'')~\cite{FermiPulsar} and the $\gamma$/hadron separation cuts, we end up with 58 hours of effective time for the analysis.

\subsection{Morphology}

\begin{figure}[!t]
  \vspace{5mm}
  \centering
  \includegraphics[width=3.2in]{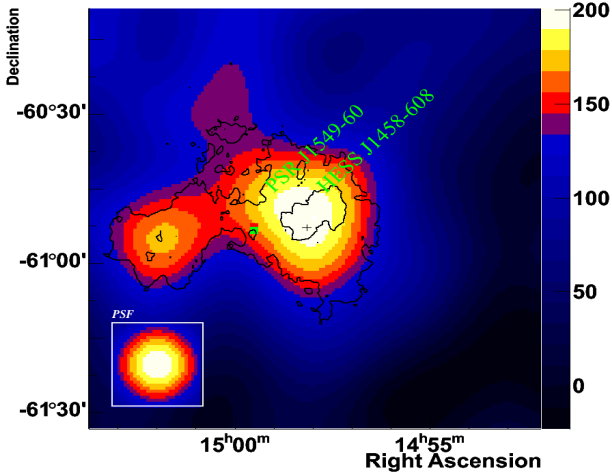}
  \caption{VHE gamma-rays excess map of HESS J1458-608. The excess counts have been smoothed with a Gaussian of width 13.2'. The blue-red transition in the color scale is set at approximately 3$\sigma$ significance. The black contours correspond to 5 and 7$\sigma$ significance. The position of the pulsar PSR J1459-60 is marked by a green and red filled square, which correspond to {\it Fermi}/LAT and {\it Suzaku} best fit position respectively. The best fit position of HESS J1458-608 is marked by a black cross and the error bars correspond to 1$\sigma$ error in the position fit.}
  \label{excessskymap}
\end{figure}

A fit of a 2-D model of a symmetrical gaussian, convolved with the telescope point spread function (PSF), to the uncorrelated image gives the centroid of the source (black cross in figure~\ref{excessskymap}) at $\alpha_{J2000}$=14$^h$58$^m$9.6$^s$ ($\pm$ 7$^s$) and $\delta_{J2000}$=-60$^\circ$52'38'' ($\pm$ 4''). The 1$\sigma$ extension of the fitted Gaussian is 0.17$^\circ$ $\pm$ 0.07$^\circ$. 
The best fit position locates HESS J1458-608 at $\sim$ 0.16$^\circ$ to the West of the {\it Fermi} pulsar position.
Figure\ref{excessskymap} shows the raw excess counts sky map smoothed with a symmetrical Gaussian with a standard deviation of 13.2'. The best fit position is marked as a black cross.
A 2-D model of a combination of two symmetrical gaussians has been rejected at a level of $<$ 3$\sigma$.

An analysis at the best fit position of HESS J1458-608 with a cut $\theta$ = 0.33$^\circ$ ($\sim$ 95\% of source enclosure) gives a significance of 8.1$\sigma$ ($\sim$ 6$\sigma$ post-trials) and reveals a clear extended emission around the pulsar position with an excess maximum offset by $\sim$10 arcmin from the pulsar position (Fig.~\ref{excessskymap}).

\subsection{Spectrum}

The results of the spectral analysis for HESS J1458-608 presented here were obtained using a circular extraction region centered in the best fit position, with a radius corresponding to the 95\% source enclosure ( $\theta$=0.33$^\circ$). 

\begin{figure}[!t]
  \vspace{5mm}
  \centering
  \includegraphics[width=3.2in]{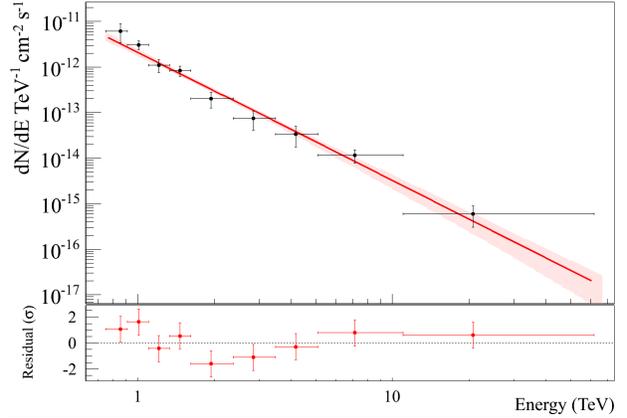}
  \caption{Energy differential spectrum of HESS J1458-608 at the best fit position. The resulting power law fit, between 0.750 and 60 TeV, is displayed as a solid red line. The plot at the bottom shows the residuals of the power-law fit to the data points (black filled dots in the top panel).}
  \label{spectrum}
\end{figure}

Figure~\ref{spectrum} shows the photon spectrum, well fitted by a power law (dN/dE = K$\cdot$E$^{-\Gamma}$), displayed as a red line in the same figure.
The spectral fit used a forward folding algorithm with a minimum significance per bin of 2$\sigma$. The fit results on a photon index of $\Gamma$= 2.8 $\pm$ 0.2 and differential flux normalization at 1 TeV K = (2.1 $\pm$ 0.4)$\times$10$^{-12}$ TeV$^{-1}$cm$^{-2}$s$^{-1}$. 
The corresponding integral flux above 1 TeV is (1.2 $\pm$ 0.4)$\times$10$^{-12}$ cm$^{-2}$s$^{-1}$, which corresponds to $\sim$6\% of the Crab Nebula flux at the same energies.


\section{Possible associations}

An extensive search on the archival data at other wavelengths has not revealed any counterpart at the coordinates of HESS J1458-608.

The only counterpart in the vicinity of HESS J1458-608 is the {\it Fermi} pulsar PSR J1459-60 located within the extended emission at 9.6' from the source fit position.

The pulsar PSR J1459-60 ($\alpha_{J2000}$=14$^h$59$^m$29.8$^s$, $\delta_{J2000}$=-60$^\circ$52'40'')~\cite{FermiPulsar} was discovered in $\gamma$-rays, followed by the detection of a faint counterpart in X-rays~\cite{X-raythesis}~\cite{x-raySwift}. No radio counterpart has been reported up to now.
The pulsar has a period of 103 ms, a surface magnetic field B=1.64$\times$10$^{12}$G, a spin-down age of $\tau_c$=64 kyr and a spin-down energy loss rate 9.2$\times$10$^{35}$ erg/s. 
The fitted spectrum to the pulsed emission follows a power-law function plus an exponential cut-off, represented by $dN/dE=K \cdot E^{-\Gamma} exp(-E/E_c)$ with $\Gamma$=1.83$\pm$0.24 and an exponential cut-off E$_c$=2.7$\pm$1.1 GeV. The integral energy flux above 100 MeV is F($>$100 MeV) = 10.6$\pm$1.2$\times$10$^{-11}$ erg cm$^{-2}$ s$^{-1}$\cite{FermiPulsar}.

To find the corresponding PWN contribution from the {\it Fermi} pulsars, {\it Fermi} collaboration has carried out an analysis in the off-pulsed emission region for all the $\gamma$-ray pulsars. No significant emission has been found in the case of PSR J1459-60. Only 4 PWNe have been detected with 1-yr sensitivity of {\it Fermi}~\cite{FermiPWN}, all of them associated with young (1-10 kyr) and bright (log$\dot{E}$$>$ 36.85 erg/s) pulsars. Clearly PSR J1459-60 is beyond those limits.

X-rays observations by {\it Suzaku}~\cite{X-raythesis} have confirmed point-like non-thermal emission~\cite{X-raythesis} at 0.8 arcmin from the $\gamma$-ray pulsar position. The emission was fitted to two models: an absorbed power-law $dN/dE=exp[-N_H\cdot\sigma(E)]\times K_{PL}(-\Gamma+2)E^{-\Gamma}/(E_2^{-\Gamma+2}-E_1^{-\Gamma})$ and an absorbed power-law plus black-body. Black-body contribution was ruled out from the fit and the absence of of emission/absorption lines in the spectra. Therefore the best fit model to the X-ray emission is an absorbed power-law with $\Gamma$=1.54$^{+0.39}$$_{-0.30}$, K$_{PL}$=(0.142$\pm$0.02)$\times$10$^{-12}$erg cm$^{-2}$ s$^{-1}$. The estimated distance from this analysis is 4 kpc.

Recently, pulsed emission in X-rays has been reported by Swift~\cite{x-raySwift} at an offset of 9.8'' from Fermi ephemeris.

\section{Discussion}

The lack of an enhancement of the molecular cloud content and the location of the Fermi pulsar within the HESS J1458-608 extended emission suggests a leptonic scenario as the origin of the TeV emission. 

Despite the offset on the position of the pulsar and HESS J1458-608, the hypothesis that HESS J1458-608 is the pulsar wind nebula of the old $\gamma$-ray pulsar PSR J1459-60 seems likely.

\begin{figure}[!t]
  \vspace{5mm}
  \centering
  \includegraphics[width=3.0in]{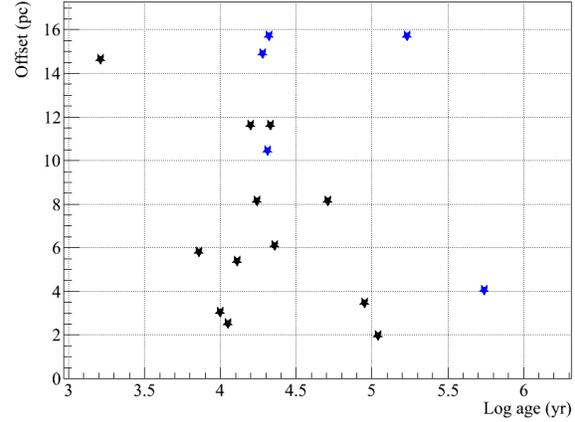}
  \caption{Offset (in pc) between the X-ray and the TeV emission for a population of pulsar-PWN systems~\cite{Kargaltsev} as a function of the pulsar age (spin-down age). The black points correspond to those sources with X-ray emission detected while the blue points to those which X-ray emission is faint or nonexistent. HESS J1458-608 - PSR J1459-60 system have an offset of 11 pc and a log($\tau_c$)=4.8.}
  \label{PWNpopulation}
\end{figure}

From the morphological point of view and considering the the pulsar and its PWN are at the same distance to the observer, the fact that the pulsar is displaced from the bulk of the TeV emission can be explained by the kick-off momentum transmitted to the pulsar on its birth as a result of a non symmetric supernova explosion. 
Considering the age of PSR J1459-60 (64 kyr), the estimated distance (4 kpc) and the offset between the current pulsar position and the peak of TeV emission (9.6 arcmin) one can estimate the pulsar velocity $\sim$200 km/s. This velocity falls into the range of the typical pulsar velocities~\cite{Romani}. Therefore, the pulsar could have moved from the birth location by $\sim$11 (d/4 kpc) pc during its lifetime. In fact, evolved PWN systems typicaly surround old pulsars, displaced few arc minutes from the PWN center~\cite{deJager2008}. Figure~\ref{PWNpopulation} shows the offset between the pulsar and the TeV emission of the associated PWN as a function of the pulsar characteristic age $\tau_c$~\cite{Kargaltsev}. For old pulsar-PWN systems (log($\tau_c$) (yr) $>$ 4) the typical offset values seem to be between $\sim$ 2-16 pc.

This scenario of the evolved PSR-PWN system described above is consistent with the spectral results from X-rays and TeV $\gamma$-rays.
The electrons producing the synchrotron emission (X-rays between 0.5-10 keV) at the pulsar position would require interstellar magnetic fields of the order of 0.4-1.5$\mu$G, much lower than the typical minimum value of $\sim$ 3-5$\mu$G~\cite{Aharonian1997}, if they were produced 64 kyr ago. Therefore, the faint X-ray emission is produced by a recently emitted population of electrons.
On the other hand, the electrons responsible for the TeV might be much older, corresponding to a population of electrons emitted by the pulsar at its birth (64 kyr ago). The IC emission of such electrons interacting with the CMB would produce the VHE gamma-rays from HESS J1458-608 at energies below 30 TeV \cite{Aharonian1997} (presented in Fig.~\ref{spectrum}). This old electron population, below 100 TeV, would have a spectral index of 1.8, which is typical of PWN scenarios.

Assuming that HESS J1458-608 is the PWN of the pulsar PSR J1459-60 we can estimate the conversion efficiency between the spin-down energy ($\dot{E}$) and the TeV luminosity above 1 TeV $\eta_{TeV}$=L$_{TeV}$/$\dot{E}$=2\% while the conversion efficiency in X-rays is $\eta_X$=0.01\%. These two values give a luminosity ratio TeV/keV (log L$_{TeV}$/L$_x$) of the order of 2, which is within the range between 1 and 3 typical of old PWN systems~\cite{Kargaltsev}.

\section{Summary}

The newly discovered source HESS J1458-608 has been detected at VHE by H.E.S.S. and its extension includes the energetic {\it Fermi}/LAT pulsar PSR J1459-60, which faint emission in X-rays have been recently confirmed by {\it Suzaku} and {\it Swift}.
During the search of counterparts in the HESS J1458-608 FoV no other significant counterpart at other wavelengths has been found at the best fit position of the TeV source. 

A morphological study of the TeV emission and the pulsar PSR J1459-60 seems to link the two objects into an evolved PSR-PWN scenario, where the pulsar has moved by $\sim$11 pc from its birth place, located at the best-fit position of HESS J1458-608.
A leptonic model could explain the synchrotron emission produced by a recent population of electrons at the pulsar position while the IC from older population of electrons would be located at 9.6 arcmin to the West of PSR J1459-60.
Within this scenario, the aging of the pulsar might also explain the lack (or faint) of X-ray emission closer to the TeV emission, although dedicated observations at the HESS J1458-608 position would be needed to confirm the first results' hypothesis.

\section{Acknowledgments}

The support of the Namibian authorities and of the University of Namibia in facilitating the construction and operation of H.E.S.S. is gratefully acknowledged, as is the support by the German Ministry for Education and Research (BMBF), the Max Planck Society, the French Ministry of Research, the CNRS-IN2P3 and the Astroparticle Interdisciplinary Programme of the CNRS, the U.K. Science and Technology Facilities Council (STFC), the IPNP of the Charles University, the Polish Ministry of Science and Higher Education, the South African Department of Science and Technology and National Research Foundation, and by the University of Namibia. We appreciate the excellent work of the technical support staff in Berlin, Durham, Hamburg, Heidelberg, Palaiseau, Paris, Saclay, and in Namibia in the construction and operation of the equipment. AZ acknowledges partial support from MNiSW grant N203 387737.


\clearpage

\end{document}